\documentclass[aps,pra,showpacs,amsmath,amssymb,lengthcheck]{revtex4}

\usepackage{epsfig}

\begin{document}

\title{Normal-Superfluid Interface for Polarized Fermion Gases}

\author{{\small Bert Van Schaeybroeck and Achilleas Lazarides}}
\affiliation{ Instituut voor Theoretische Fysica, Katholieke
Universiteit
  Leuven,\\ Celestijnenlaan 200 D, B-3001 Leuven, Belgium.}

\date{\small\it \today}

\begin{abstract}
Recent experiments on imbalanced fermion gases have proved the
existence of a sharp interface between a superfluid and a normal
phase. We show that, at the lowest experimental temperatures, a
temperature difference between N and SF phase can appear as a
consequence of the blocking of energy transfer across the
interface. Such blocking is a consequence of the existence of a SF
gap, which causes low-energy normal particles to be reflected from
the N-SF interface. Our quantitative analysis is based on the
Hartree-Fock-Bogoliubov-de Gennes formalism, which allows us to
give analytical expressions for the thermodynamic properties and
characterize the possible interface scattering regimes, including
the case of unequal masses. Our central result is that the thermal
conductivity is exponentially small at the lowest experimental
temperatures.
\end{abstract}


\maketitle

\section{Introduction}
By mixing fermion species of unequal particle number, recent
experiments have revealed many aspects of the exciting physics of
imbalanced fermion
gases~\cite{partridge,partridge2,zwierlein1,zwierlein2,zwierlein3,zwierlein4,zwierlein5}.
In these experiments, a trap is loaded with an ultracold spin
mixture consisting of two lithium hyperfine states. The existence
of an easily-accessible Feshbach resonance allows the tuning of
the interspecies interactions with great accuracy so that the
entire BEC-BCS crossover with number imbalance may be
explored~\cite{houbiers}. The \textit{BCS regime} involves a small
and negative s-wave scattering length, with the fermions forming
weakly bound Cooper pairs. On the other hand, in the \textit{BEC
regime}, where the scattering length is positive, the particles
form tightly bound molecular pairs forming a
condensate~\cite{houbiers,sheehy2,giorgini,ketterle}. In between
lies the \textit{unitary regime}, characterized by a diverging
scattering length (signifying the appearance of a bound state),
where particles are strongly interacting. The divergence of the
scattering length implies that the system is universal, in the
sense that only a single length scale characterizes it.

Much work has been done in an attempt to understand the ground
state properties of the imbalanced clouds; however, despite
theoretical
advances~\cite{sheehy,chen,chevy,gubbels,gubbels2,bedaque,parish,desilva,haque,lazarides,lobo,klimin},
some problems remain unresolved. The MIT
data~\cite{zwierlein1,zwierlein2,zwierlein3,zwierlein4,zwierlein5}
are in good agreement with Monte Carlo (MC)
simulations~\cite{lobo} and renormalization group (RG)
predictions~\cite{gubbels}, which point to the existence of a
\textit{three-shell} structure consisting of a superfluid (SF)
core phase, a partially polarized and a fully-polarized normal (N)
phase. From these experiments, the homogeneous phase diagram has
been determined and is in agreement with both the MC and RG
results. On the other hand, the results of the Rice group's
experiments~\cite{partridge,partridge2} are are less well
understood. For example, they find that superfluidity does not
break down at high polarizations, which is incompatible with the
results of the MIT experiments and thus also in contradiction to
theoretical predictions of a critical polarization, above which
superfluidity breaks down. Since both the MIT phase diagram and
the MC and RG predictions were obtained under the assumption that
the local density approximation (LDA) holds, and since the Rice
experiments involve smaller traps, it is likely that the answer to
the riddle lies in the breakdown of LDA (in other words,
finite-size effects); indeed, taking as an ansatz a
\textit{two-shell} structure, the Rice data can be recovered by
using a N-SF interface tension as a fitting
parameter~\cite{desilva,haque}.

For our purposes, it is sufficient to say that the MIT experiments
are well-described by theory, and furnish direct evidence that a
N-SF interface exists, separating an unpolarized SF from a
partially polarized N.

In this article we argue that the N-SF interface may block the
thermal equilibration process between the N and SF phases; one
consequence of this will be to cause a temperature imbalance to
appear across the interface. Incorporating such temperature
difference in the existing models may provide the key to a better
understanding of the experiments. We present a detailed study of
the scattering off the N-SF interface within the
Hartree-Fock-Bogoliubov-de Gennes (HFBdG) formalism, followed by a
calculation of the heat conductivity across the interface. We find
that this conductivity drops exponentially fast with increasing
$\Delta/k_{_B} T$ with $\Delta$ the SF gap and $T$ the
temperature. Our model predicts the conductivity to be vanishingly
small for the experiments at unitarity. A summary of our results
was published earlier in Ref.~\onlinecite{vanschaeybroeck}.

The article is structured as follows. We start off in
section~\ref{sec_bulk} by discussing the bulk physics of
imbalanced fermion gases. This is done using a mean-field
Hamiltonian which incorporates both the gap formation and the
Hartree-Fock terms. We present analytical results for the
equations of state which enable us to write down the condition for
coexistence between the N and the SF phases. Then, in
section~\ref{sec_intscatt}, we discuss in detail how particles
scatter on a N-SF interface according to the
Hartree-Fock-Bogoliubov-de Gennes formalism. We find a rich
variety of interfacial scattering processes, depending on the
energy and momentum of the incident particle. The implications of
this scattering are outlined in section~\ref{sec_transort}, where
we obtain the thermal conductivity exactly and with a good
analytical approximation. Finally in section~\ref{sec_timescales}
we calculate that for the experiments at lowest temperatures, the
typical timescale associated with the equilibration of a
temperature difference across the interface, is of order seconds
and therefore comparable with the trap lifetime.

\section{Bulk Physics and Coexistence}\label{sec_bulk}
\subsection{Bulk Phases}
Fermion mixtures in the BEC-BCS crossover are the subject of a
very active research field. Although some theoretical models give
accurate results for the BEC and/or BCS regimes, no model yet
allows an analytical, quantitative description across the entire
crossover.
In the following, we use the so-called \textit{one-channel model}
at zero temperature; this is a straightforward generalization of
the BCS model to arbitrarily strong coupling and is widely used in
the literature, as it is known to embody most qualitative features
of the BEC-BCS crossover for the currently experimentally relevant
Feshbach resonances. Interactions between fermions and the bosonic
bound states may be explicitly incorporated within a two-channel
model~\cite{sheehy2}.

We start off by deriving the Hartree-Fock-Bogoliubov-de Gennes
(HFBdG) equations. We then focus on the physics of an interface
separating two semi-infinite phases: a SF and a polarized N. We
write down the equations of state in these phases, and use them to
find the condition for two-phase coexistence. In this section, we
extend the work reported in Ref.~\onlinecite{vanschaeybroeck} by
including the HF term into the BdG equations and treating the case
of unequal species masses.

The system under consideration consists of two fermionic species
$\uparrow$ and $\downarrow$ with masses $m_i$ and chemical
potentials $\mu_i$ for $i=\uparrow,\,\downarrow$ and which we
assume to be trapped by the external potential $V_i(\mathbf{r})$.
The intraspecies interactions are via p-wave scattering which can
be safely ignored at low temperatures. On the other hand, the
interactions between the two spin species are taken to be contact
interactions characterized by the coupling constant $G=4\pi
\hslash^2 a/m_{_+}<0$ with $m_{_+}\equiv2(m_\uparrow^{-1}+
m_\downarrow^{-1})^{-1}$ and $a$ the s-wave scattering length.
Within a generalized HF approximation, one may work with the
following Hamiltonian~\cite{degennes,fetter}:
\begin{align}\label{diagonl}
\widehat{H}=\int_{_V}\text{d}\mathbf{r}\,&\left[
\widehat{\Psi}_\uparrow^{\dag}(\mathbf{r})\mathcal{H}_\uparrow(\mathbf{r})\widehat{\Psi}_\uparrow(\mathbf{r})
+\widehat{\Psi}_\downarrow^{\dag}(\mathbf{r})\mathcal{H}_\downarrow(\mathbf{r})\widehat{\Psi}_\downarrow(\mathbf{r})\right.\\
&\quad\left.+\Delta(\mathbf{r})\widehat{\Psi}_\uparrow^{\dag}(\mathbf{r})\widehat{\Psi}_\downarrow^{\dag}(\mathbf{r})
+\Delta^{*}(\mathbf{r})\widehat{\Psi}_\downarrow(\mathbf{r})\widehat{\Psi}_\uparrow(\mathbf{r})\right],\nonumber
\end{align}
where
$\mathcal{H}_i\equiv-\hslash^{2}\boldsymbol{\nabla}^{2}/(2m_i)-\mu_i+V_i(\mathbf{r})+U_i(\mathbf{r})$,
the gap
$\Delta=-G\langle\widehat{\Psi}_\uparrow(\mathbf{r})\widehat{\Psi}_\downarrow(\mathbf{r})\rangle$
and the HF terms
$U_\uparrow(\mathbf{r})=G\langle\widehat{\Psi}_\downarrow^{\dag}(\mathbf{r})\widehat{\Psi}_\downarrow(\mathbf{r})\rangle$
and
$U_\downarrow(\mathbf{r})=G\langle\widehat{\Psi}_\uparrow^{\dag}(\mathbf{r})\widehat{\Psi}_\uparrow(\mathbf{r})\rangle$.
As usual, one diagonalizes the Hamiltonian~\eqref{diagonl} by a
Bogoliubov transformation i.e.~by writing the operators $\Psi_i$
in terms of fermionic quasiparticle operators $\hat{c}$:
\begin{align*}
\widehat{\Psi}_\uparrow(\textbf{r},t)&=\sum_{\ell} e^{-i E_{\ell}
t/\hslash} \left[u_{\ell ,\uparrow}(\textbf{r})\hat{c}_{\ell
,\uparrow}-v_{\ell ,\uparrow}(\textbf{r})\hat{c}^\dag_{\ell,\downarrow} \right],\\
\widehat{\Psi}_\downarrow(\textbf{r},t)&=\sum_{\ell} e^{-iE_{\ell}
t/\hslash} \left[u_{\ell ,\downarrow}(\textbf{r})\hat{c}_{\ell
,\downarrow}-v_{\ell,\downarrow}(\textbf{r})\hat{c}^\dag_{\ell,\uparrow}
\right],
\end{align*}
where $u_{\ell,i}$ and $v_{\ell,i}$ are the quasiparticle wave
functions, $\{\hat{c}_{\ell,i},\hat{c}_{\ell',j}\}=0$ and
$\{\hat{c}_{\ell
,i}^{\dag},\hat{c}_{\ell',j}\}=\delta_{\ell,\ell'}\delta_{i,j}$.
Then, if under this transformation the Hamiltonian~\eqref{diagonl}
is diagonalized, the relations
$[\widehat{H},\,\hat{c}_{\ell,i}]=-E_\ell \hat{c}_{\ell,i}$ and
$[\widehat{H},\,\hat{c}^{\dag}_{\ell,i}]=E_n
\hat{c}^{\dag}_{\ell,i}$ must hold such that one straightforwardly
finds the four \textit{Hartree-Fock-Bogoliubov-de Gennes} (HFBdG),
or \textit{Blonder-Tinkham-Klapwijk},
equations~\cite{degennes,BTK}:
\begin{subequations}\label{Bdg}
\begin{align}
\mathcal{H}_\uparrow u_{\ell,\uparrow}+\Delta v_{\ell,\downarrow}&=E_\ell u_{\ell,\uparrow},\\
\Delta^* u_{\ell,\uparrow}-\mathcal{H}_\downarrow v_{\ell,\downarrow}&=E_\ell v_{\ell,\downarrow},\\
\mathcal{H}_\downarrow u_{\ell,\downarrow}+\Delta v_{\ell,\uparrow}&=E_\ell u_{\ell,\downarrow},\\
\Delta^* u_{\ell,\downarrow}-\mathcal{H}_\uparrow
v_{\ell,\uparrow}&=E_\ell v_{\ell,\uparrow}.
\end{align}
\end{subequations}
The wave functions $u_{\ell,\uparrow}$ and $v_{\ell,\downarrow}$
which appear in the first two equations are not coupled to the
wave functions $u_{\ell,\downarrow}$ and $v_{\ell,\uparrow}$ of
the last two equations so that, apart from normalization, they can
be treated separately.

In the rest of this section, we will be concerned with the study
of a homogeneous system of two fermion species at fixed chemical
potentials $\mu_\uparrow$ and $\mu_\downarrow$ and with no
external potential $V_\uparrow=V_\downarrow=0$. Furthermore we
treat both the case of a N and a SF phase.

At zero temperature and for negative scattering length, there
always exists a (meta-)stable state with $\Delta=\Delta^{*}\neq 0$
and for which both species densities are equal; we call this state
the \textit{BCS state} or \textit{SF phase}. For this state, we
define now the \textit{effective chemical potential imbalance}
$h_{_S}$ and the \textit{average effective chemical potential}
$\mu_{_S}$ as follows:
\begin{subequations}
\begin{align}\label{chems}
\mu_{_S}&\equiv\frac{\mu_\uparrow+\mu_\downarrow}{2}-U_{_S},\\
h_{_S}&\equiv\frac{\mu_\uparrow-\mu_\downarrow}{2}.
\end{align}
\end{subequations}
Without loss of generality, we further assume that $h_{_S}>0$.
Note that in the SF phase, both spin species have equal densities
which implies equal HF potentials $U_{_S}=U_\uparrow=U_\downarrow$
for the spin species. One can then find the ground state energy of
the \textit{homogeneous} SF system as a function of
$\mu_\uparrow$, $\mu_\downarrow$ and $a$. Solving the HFBdG
Eqs.~\eqref{Bdg}, and minimizing the ground state energy with
respect to the variables $\Delta$ and $U_{_S}$, the gap and number
equations at zero temperature are~\footnote{Due to the
momentum-independent interaction strength, one must cancel the
ultraviolet divergencies originally appearing in the integral. In
the presence of a HF term this was treated in
Refs.~\cite{ohashi,randeria,bulgac2,houbiers,tempere,
papenbrock,su,grasso} for homogeneous systems.}:
\begin{subequations}
\begin{align}
1&=-\frac{G}{2}\int
\frac{\text{d}^3\mathbf{k}}{(2\pi)^3}\left[\frac{1}{\sqrt{(\varepsilon_{_\mathbf{k}}-\mu_{_S})^2+\Delta^2}}
-\frac{1}{\varepsilon_{_\mathbf{k}}}\right],\\
U_{_S}&=-\frac{G}{2}\int
\frac{\text{d}^3\mathbf{k}}{(2\pi)^3}\left[1-\frac{\varepsilon_{_\mathbf{k}}-\mu_{_S}}
{\sqrt{(\varepsilon_{_\mathbf{k}}-\mu_{_S})^2+\Delta^2}}\right],
\end{align}
\end{subequations}
where
$\varepsilon_{_\mathbf{k}}\equiv\hslash^2\mathbf{k}^2/(2m_{_+})$.
Note that the above equations are coupled by the term $U_{_S}$
which appears in the definition of $\mu_{_S}$. The ground state
grand potential per unit volume in terms of $\Delta$ and
$\mu_{_S}$ is:
\begin{align}\label{grapot}
\Omega_{_S}&=\int
\frac{\text{d}^3\mathbf{k}}{(2\pi)^3}\left[\varepsilon_{_\mathbf{k}}-\mu_{_S}
-\frac{\Delta^2/2+(\varepsilon_{_\mathbf{k}}-\mu_{_S})^2}{\sqrt{(\varepsilon_{_\mathbf{k}}-\mu_{_S})^2+\Delta^2}}\right].
\end{align}
We are mostly interested in the BCS side of the BEC-BCS crossover
where the effective chemical potentials
$\mu_{_i}-\mathrm{U}_{_{i}}$ and $\mu_{_S}$ can be taken as
positive. Performing the integrals over the three-dimensional wave
vectors leads to the analytical expressions for the gap, the HF
potential and the grand potential per unit
volume~\cite{gradshteyn,papenbrock,su,marini}:
\begin{subequations}\label{eqstate}
\begin{align}
\epsilon_{_a}&=-2\mu_{_S}[P_{_{1/2}}(\eta)]^2/\eta,\label{eqstate2}\\
U_{_S}
&=\mu_{_S}\left[1-\eta^{-1}P_{_{3/2}}(\eta)/P_{_{1/2}}(\eta)\right],\label{eqstate3}\\
\Omega_{_S}
&=-\frac{(m_{_+})^{3/2}\epsilon_{_a}^{5/2}\eta^2 }{80\pi\hslash^3|P_{_{1/2}}(\eta)|^5}\nonumber\\
&\quad\times\left[(\eta^{-2}-5)P_{_{1/2}}(\eta)+4\eta^{-1}P_{_{3/2}}(\eta)\right].\label{eqstate1b}
\end{align}
\end{subequations}
with $P$ the Legendre function,
$\eta\equiv-1/\sqrt{1+(\Delta/\mu_{_S})^2}$\label{zeradef} and we
introduced the energy scale which is set by the scattering length
$\epsilon_{_a}\equiv\hbar^2/(m_{_+}a^2)$\label{scattdef}.
Expression~\eqref{eqstate1b} allows to write the grand potential
in terms of $\Delta/\epsilon_{_a}$; indeed, for each value
thereof, one can find $\eta$ by solving the relation:
\begin{align}\label{vergelijk}
1=\eta^{2}+4(\Delta/\epsilon_{_a})^2[P_{_{1/2}}(\eta)]^4,
\end{align}
such that the grand potential~\eqref{eqstate1b} is fully
determined by $\Delta/\epsilon_{_a}$ only.

For the \textit{normal state}, one finds the following expressions
for the grand potential per unit volume of the ground state and
for the HF potentials:
\begin{subequations}\label{grootpot}
\begin{align}
\Omega_{_N}&=-\frac{2\sqrt{2}}{15\pi^2\hslash^3}\left[m_\uparrow^{3/2}(\mu_\uparrow-U_\uparrow)^{5/2}\right.\label{normeqstate1}\\
&\quad\quad\quad\quad\quad\quad\quad\quad\quad\quad\left.+m_\downarrow^{3/2}(\mu_\downarrow-U_\downarrow)^{5/2}\right]\nonumber,\\
\frac{U_\uparrow}{\epsilon_{_a}}&=-\frac{4\sqrt{2}}{3\pi}\left(\frac{m_\downarrow}{m_{_+}}\right)^{3/2}\left(\frac{\mu_\downarrow-U_\downarrow}{\epsilon_{_a}}\right)^{3/2}\label{normeqstate2},\\
\frac{U_\downarrow}{\epsilon_{_a}}&=-\frac{4\sqrt{2}}{3\pi}\left(\frac{m_\uparrow}{m_{_+}}\right)^{3/2}\left(\frac{\mu_\uparrow-U_\uparrow}{\epsilon_{_a}}\right)^{3/2}\label{normeqstate3}.
\end{align}
\end{subequations}
At fixed chemical potentials, the attractive interactions induce
an increased normal state density and a decreased ground-state
grand potential. Eqs.~\eqref{normeqstate2}
and~\eqref{normeqstate3} are coupled in a non-trivial manner and
can be solved numerically to find $U_\uparrow$ and $U_\downarrow$.

\subsection{Bulk Two-Phase Coexistence}
Before writing down the N-SF coexistence condition, we explain how
a variation of the chemical potential imbalance between the spin
species may change the ground state. A small imbalance (or
polarization) gives rise to a SF state while a large polarization
leads to a N state. Intuitively this is expected since it is known
from BCS theory that pair formation is energetically favorable for
particles with opposite momenta. By shifting the Fermi energies,
an energy cost must be paid which is associated with the matching
of the Fermi levels so as to enable pair formation; when this
energy cost is too high, the N state prevails.

In a \textit{trapped} imbalanced fermion system, both the N and
the SF phase may be present, bounded by an interface. Using a
sufficiently large number of particles, a local density
approximation is justified there and amounts to using an effective
local chemical potential
$\mu_i(\mathbf{r})=\mu^0_i-V_i(\mathbf{r})$ with $\mu^0_i$ the
chemical potential at the trap center and $i=\uparrow,\,
\downarrow$~\cite{bulgac,
haque,bertaina,haque2,haussman,recati,zwierlein3}. Since in
experiments both spin species feel the same trapping potential,
the imbalance
$\mu_\uparrow(\mathbf{r})-\mu_\downarrow(\mathbf{r})$ is constant
throughout the trap. On the other hand, the average chemical
potential
$[\mu_\uparrow(\mathbf{r})+\mu_\downarrow(\mathbf{r})]/2$ varies:
it is large at the center of the trap and decreases towards the
trap boundary. The ground state energy of the SF increases with
increasing value of the gap; as seen from Eqs.~\eqref{eqstate2},
the gap value is proportional to the average chemical potential
$\mu_{_S}$ and independent of $h_{_S}$. Therefore, if present, the
SF phase will be at the trap center where $\mu_{_S}$ is maximal.
Moving radially outwards, two-phase coexistence of the N and SF
phase occurs at the position where there is a balance between the
energy gained by creating, on the one hand, a gap in the SF, and,
on the other hand, a density difference in the N phase.

To establish now the condition for coexistence between the N and
the SF phase, recall that we have a system containing two fermion
species $i=\uparrow,\, \downarrow$ with masses $m_i$ and chemical
potentials $\mu_i$ for which interactions are characterized by a
negative scattering length $a$. We aim to find the coexistence
condition as a function of the parameters $a$, $\mu_\uparrow$ and
$\mu_\downarrow$. The SF state is fully described by the
parameters $\Delta$ and $U_{_S}$, which can be extracted from
Eqs.~\eqref{eqstate2},~\eqref{eqstate3} and~\eqref{vergelijk}. The
normal state, on the other hand, is characterized by $U_\uparrow$
and $U_\downarrow$ which are solved for using
Eqs.~\eqref{normeqstate2} and~\eqref{normeqstate3}.

The coexistence condition obtained by equating the grand
potentials of Eqs.~\eqref{eqstate1b} and~\eqref{normeqstate1} is:
\begin{widetext}
\begin{align}\label{coexistence}
\frac{3\sqrt{2}\pi\eta^2}{64|P_{_{1/2}}(\eta)|^5}\left[(\eta^{-2}-5)P_{_{1/2}}(\eta)+4\eta^{-1}P_{_{3/2}}(\eta)\right]=\left(\frac{m_\uparrow}{m_{_+}}\right)^{3/2}\left(\frac{\mu_\uparrow-U_\uparrow}{\epsilon_{_a}}\right)^{5/2}
+\left(\frac{m_\downarrow}{m_{_+}}\right)^{3/2}\left(\frac{\mu_\downarrow-U_\downarrow}{\epsilon_{_a}}\right)^{5/2}.
\end{align}
\end{widetext}
By condition~\eqref{coexistence} one can find the surface of
coexistence in the space of the parameters
($\mu_\uparrow,\mu_\downarrow,a$).

For one specific case it is easy to relate the gap value to the
chemical potentials at coexistence; this occurs in case of equal
masses $m_{_+}=m_\uparrow=m_\downarrow$ without incorporation of
the HF term. The energy scale $\epsilon_{_a}$ can then be
eliminated and by a numerical analysis, one finds to a good
approximation the Clogston relation, originally derived for
superconductors in a magnetic field~\cite{clogston}:
\begin{align}\label{clogston}
 \Delta=\sqrt{2}h_{_S}.
 \end{align}
The Clogston limit is exact in the BCS limit, when $h_{_S}\ll
\mu_{_S}$.

Lastly, it is common to write the scattering length $a$ in units
of $1/k_{_F}$, the latter defined by the relations
$E_{_F}=k_{_B}T_{_F}=\hslash^2k_{_F}^2/(2m_{_+})=\hslash^2[3\pi^2n_{_S}]^{2/3}/(2m_{_+})$;
we find that in the SF:
\begin{align*}
k_{_F}
a=&\left[\frac{3\pi}{4}\frac{\left(P_{_{3/2}}(\eta)/P_{_{1/2}}(\eta)-\eta\right)}{P_{_{1/2}}^2(\eta)}\right]^{1/3}.\nonumber
\end{align*}

\section{Interface Scattering}\label{sec_intscatt}

\subsection{Bulk Particles}
In this section, we explain in full detail how particles scatter
from a N-SF interface for imbalanced fermion gases. By taking a
simple model for the interface, we describe the different
scattering regimes and we give the transmission coefficients for
scattering across the interface; the latter will be used for
calculating transport quantities in the next section.

To model the N-SF interface, we start with a geometry wherein the
$z=0$ plane separates the N phase at $z<0$ from the SF phase at
$z>0$ or:
\begin{align}\label{gap}
\Delta(\mathbf{r})=\Theta(z) \Delta,
\end{align}
with $\Theta$ the Heaviside function. This approximation was also
used by Andreev~\cite{andreev} for studying the
normal-superconductor interface.

A particle of energy $E$ incident on the interface may scatter to
states which have the same energy and these are deduced from the
spectra in the SF and in the N phase. In the N phase, we have for
species $i=\uparrow,\downarrow$:
\begin{align}\label{freeparticles}
E_i=\left|\frac{\hslash^2\mathbf{k}^2}{2m_i}-\mu_i+U_i\right|.
\end{align}
We depict with full lines this spectrum in the left panel of
Fig.~\ref{fig1} where one recognizes two approximately linear
energy branches $\uparrow$ and $\downarrow$ near the (unequal)
Fermi surfaces. In the SF on the other hand, one finds using the
HFBdG equations, an $\alpha$ and a $\beta$ spectrum:
\begin{align}\label{SFparticles}
E_{_{\alpha,\beta}}=\mp
h_{_S}\pm\frac{\hslash^2\mathbf{k}^2}{2m_{_-}}+\sqrt{\left(\frac{\hslash^2\mathbf{k}^2}{2m_{_+}
}-\mu_{_S}\right)^2+\Delta^2},
\end{align}
where $m_{_\pm}=2(m_\uparrow^{-1}\pm m_\downarrow^{-1})^{-1}$.
These gapped spectra are depicted in the right panel of
Fig.~\ref{fig1}. The $\alpha$ branch is composed of particles of
phase $\uparrow$ and holes of phase $\downarrow$ and vice versa
for the $\beta$ spectrum. One can find that the minimal values
attained by the $\alpha$ and $\beta$ spectra are
$2(\sqrt{\widetilde{m}}\Delta-\varepsilon_{_{\alpha,\beta}}^{_0})/(\widetilde{m}+1)$
where $\varepsilon_{_{\alpha,\beta}}^{_0}\equiv[\pm
h_{_S}(1+\widetilde{m})+\mu_{_S} (1-\widetilde{m})]/2$ with
$\widetilde{m}\equiv m_\downarrow/m_\uparrow$, whence the lowest
spectrum touches the $E=0$ axis or becomes \textit{ungapped} under
the condition that
$\sqrt{\widetilde{m}}\Delta<\varepsilon_{_{\alpha}}^{_0}$. We
assume a gapped spectrum. Since the gap function is constant in
both the N and the SF phase, we can decompose the quasiparticle
wave functions $u_\uparrow$ and $v_\downarrow$, associated with
the $\alpha$-branch, into their Fourier components:
\begin{figure}
\begin{center}
   \epsfig{figure=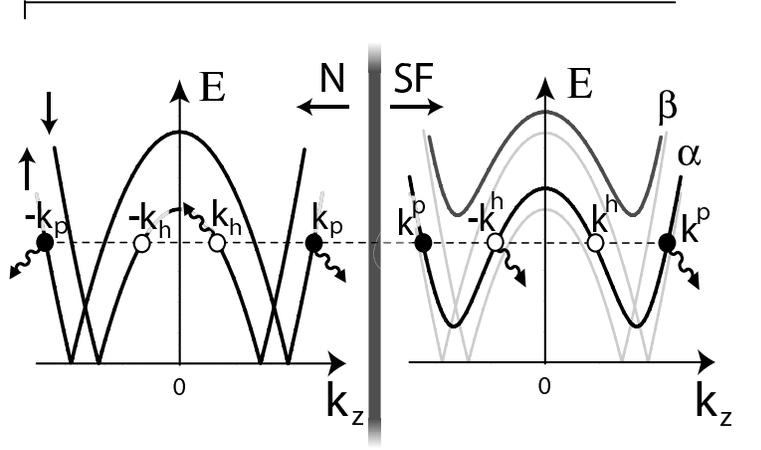,angle=0, height=140pt}
       \caption{
       The N-SF interface (thick vertical line) with the spectra of the species $\uparrow$ and $\downarrow$ on the N
       side and the gapped
       $\alpha,\,\beta$ spectra in the SF.  The long dashed line
       cuts the spectra at particle-like (filled dots) and
       hole-like (empty dots) quasiparticle states, all having the same
       energy. The minimal values of the $\alpha$ and the $\beta$
       spectrum are $2(\sqrt{\widetilde{m}}\Delta-\varepsilon_{_{\alpha,\beta}}^{_0})/(\widetilde{m}+1)$
        where $\varepsilon_{_{\alpha,\beta}}^{_0}\equiv[\pm
        h_{_S}(1+\widetilde{m})+\mu_{_S}
        (1-\widetilde{m})]/2$.
       An incident particle of species $\uparrow$ (curly arrow) with momentum $k_{_p}$ and
       energy $E$
       can have up to four scattering channels: the Andreev
       reflected $k_{_h}$ hole, the specularly reflected $-k_{_p}$ particle and
       the transmitted hole-like $-k^{_h}$ and particle-like $k^{_p}$
       states.
       \label{fig1}
     }
\end{center}
\end{figure}
\begin{align}\label{general}
 \left( \begin{array}{c}
 u_\uparrow^{_n} \\
v_\downarrow^{_n}\\
\end{array} \right)
&=\sum_{\mathbf{k},\pm} e^{i\mathbf{k}_{_\|}\cdot \mathbf{r}}
 \left( \begin{array}{c}
  A_{_{\mathbf{k},\pm}}^{_{p,n}}\,\, e^{\pm i k_{_p}z}\\
  B_{_{\mathbf{k},\pm}}^{_{h,n}}\,\, e^{\pm i k_{_h}z}
\end{array} \right),\\
\left( \begin{array}{c}
 u_\uparrow^{_s} \\
v_\downarrow^{_s}\\
\end{array} \right)
&=\sum_{\mathbf{k},\pm}e^{i\mathbf{k}_\|\cdot \mathbf{r}} \left[
\left(
\begin{array}{c}
 A_{_{\mathbf{k},\pm}}^{_{h,s}}\\
 B_{_{\mathbf{k},\pm}}^{_{h,s}}
\end{array} \right)
e^{\pm i k^{_h}z} + \left( \begin{array}{c}
  A_{_{\mathbf{k},\pm}}^{_{p,s}}\\
  B_{_{\mathbf{k},\pm}}^{_{p,s}}
\end{array} \right)
e^{\pm i k^{_p}z}\right].\nonumber
\end{align}
Here $A$ and $B$ are complex scattering amplitudes and the sub-
and superscripts $n,\, s,\, p$ and $h$ denote normal, superfluid,
particle-like and hole-like, respectively. We split the vectors
$\mathbf{k}$ into its components parallel to the wall
$\mathbf{k}_{_\|}=(k_{x},k_{y},0)$ and its $z$-component
$k_{_{p}}$, $k_{_{h}}$, $k^{_{p}}$ and $k^{_{h}}$ perpendicular to
the interface. At fixed energy, there exist eight quasiparticle
states, four of which in the N and four in the SF phase. These
states are represented by dots in Fig.~\ref{fig1}: filled dots are
particle-like states and empty dots are hole-like states (the
latter have opposite group and phase velocity). By use of the
spectra of Eqs.~\eqref{freeparticles} and~\eqref{SFparticles}, one
relates the wavenumbers $k^{_{p}}$, $k^{_{h}}$ and $k_{_h}$ to the
wavenumber $k_{_p}$ of a particle in the N phase:
\begin{subequations}\label{wavevectors}
\begin{align}
k_{_{p}}&=\left(-k_{_{\|}}^2+2m_\uparrow(\mu_\uparrow-U_\uparrow+
E_{_\alpha})/\hslash^2\right)^{1/2},\\
k_{_h}&=\left(k_{_{p}}^2+2m_\uparrow(U_\uparrow-\widetilde{m}U_\downarrow\right.\\
&\quad\quad\quad\quad\quad\quad\quad\quad\quad\left.+U_{_S}(\widetilde{m}-1)-2
\varepsilon_{_\alpha})/\hslash^2\right)^{1/2},\nonumber\\
k^{_{p,h}} &=\left(k_{_{p}}^2+2m_\uparrow(U_\uparrow-U_{_S}-
\chi_{_\alpha}^{_\pm})/\hslash^2\right)^{1/2},
\end{align}
\end{subequations}
where we defined:
\begin{subequations}\label{additional}
\begin{align}
\varepsilon_{_\alpha}&\equiv
\frac{(E_{_\alpha}+h_{_S})(1+\widetilde{m})}{2}+\frac{\mu_{_S}
(1-\widetilde{m})}{2},\label{additional1}\\
\chi_{_\alpha}^{_\pm}&\equiv\varepsilon_{_\alpha}
\mp\sqrt{\varepsilon_{_\alpha}^2-\widetilde{m}\Delta^2},
\end{align}
\end{subequations}
and $\widetilde{m}\equiv m_\downarrow/m_\uparrow$.

\subsection{Scattering Diagram}
Let us now use these equations to try to understand the physics of
the scattering processes at the interface. Consider for example a
particle incident on the interface from the N side with wavenumber
$k_{_p}$ (along the $z$-axis) and energy $E_{_\alpha}$; this is
represented in Fig.~\ref{fig1} by a curly arrow. Whatever
processes occur at the interface, energy must be conserved; but
out of the seven remaining states in the N phase with the same
energy, only four have a group velocity directed away from the
interface. this implies that
$B_{_{\mathbf{k},-}}^{_{h,n}}=A^{_{p,s}}_{_{\mathbf{k},-}}
=B^{_{h,s}}_{_{\mathbf{k},+}}=0$. The particle can now undergo
either specular reflection to the state marked in Fig.~\ref{fig1}
by $-k_{_p}$, Andreev reflection to the state $k_{_h}$ or it can
penetrate the SF as a particle-like (to $k^{_p}$) or as a
hole-like quasiparticle (to $-k^{_h}$). Note that Andreev
reflection is also of importance in other ultracold systems such
as in fermionic traps without population imbalance~\cite{urban}
and in bosonic traps near a N-SF interface in the
absence~\cite{zapata} or presence~\cite{huber} of a lattice.

The scattering amplitudes depend on the energy $E_{_\alpha}$ of
the incident particle, represented by the long dashed line in
Fig.~\ref{fig1}. Consider for instance decreasing the energy of
the incident particle to a value smaller than the lowest value of
the $\alpha$-branch; since no SF states are available,
transmission to the SF is then not possible. Mathematically this
is marked by the wavenumber of one state in the $\alpha$-branch
becoming imaginary, indicating the presence of an evanescent wave
in the SF. Furthermore, we can also raise the long dashed line of
Fig.~\ref{fig1}, that is, increase the energy $E_{_\alpha}$, to a
value above the local maximum at $k_z=0$ of the spectrum of
species $\downarrow$ in the N phase; this will inhibit Andreev
reflection.

\begin{figure}
\begin{center}
   \epsfig{figure=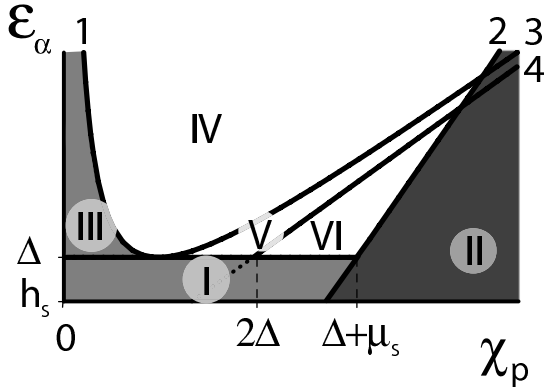,angle=0, height=140pt}
   \caption{
   The various scattering regimes for a particle of species $\uparrow$ which is incident on
     the interface from the N side, as a function of its ``energy'' $\varepsilon_{_\alpha}$ (see Eq.~\eqref{additional1})
     and
     $\chi_{_p}=\hslash^2k^2_{_p}/2m_\uparrow$ (where $k_{_p}$ is its momentum along the $z$-axis). We
     assumed here no species mass asymmetry $m_\uparrow=m_\downarrow$ and no HF potentials
     $U_{\uparrow,\downarrow,S}=0$; we depict a generic diagram for the more general
     case ($m_\uparrow\neq m_\downarrow$ and $U_{\uparrow,\downarrow,S}\neq
     0$)
     in Fig.~\ref{fig2b}. The
     heavily-shaded regimes are energetically forbidden while complete
     reflection occurs in the lightly-shaded regimes. Particles in regime~VI
     may scatter to all states indicated in Fig.~\ref{fig1} by curly arrows.
     Above line $4$ (regimes~IV and V), Andreev reflection does not occur, and
     above curve $3$ (regime~IV), hole-like excitations are also impossible. The
     functions of the curves $1$-$4$ are given in
     Eq.~\eqref{curves}.
    \label{fig2a}
    }
\end{center}
\end{figure}
\begin{figure}
\begin{center}
   \epsfig{figure=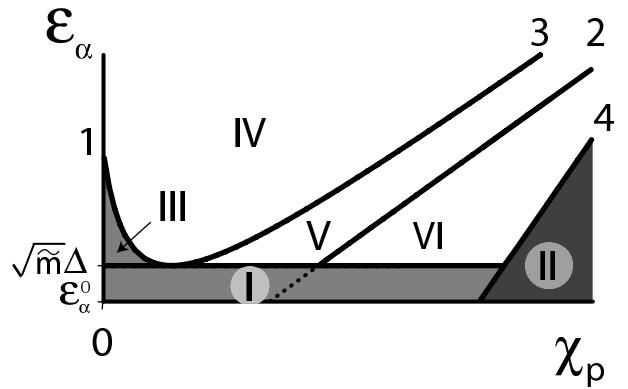 ,angle=0, height=140pt}
   \caption{The same applies as in Fig.~\ref{fig2a} except here $m_\uparrow\neq m_\downarrow$ and $U_{\uparrow,\downarrow,S}\neq
   0$. Regime I is bound by threshold line $\varepsilon_{_\alpha}=\sqrt{\widetilde{m}}\Delta$ and
     the lower limit $\varepsilon_{_\alpha}=\varepsilon_{_\alpha}^{_0}=[h_{_S}(1+\widetilde{m})+\mu_{_S}
    (1-\widetilde{m})]/2$. Compared to Fig.~\ref{fig2a}, it can be seen that curves $1$, $2$, $3$ and $4$ are shifted.\label{fig2b}
    }
\end{center}
\end{figure}

The crossover of either of the $k_{_h},\,k^{_h}$ and $k^{_p}$ and
the wave vector $\mathbf{k}_{_\|}$ from real to imaginary (or vice
versa) signifies a change in the scattering mechanism. Defining
$\chi_{_p}\equiv \hslash^2 k_{_p}^2/2m_\uparrow$\label{chipdef},
such transitions are encountered when (see
Eq.~\eqref{wavevectors}):
\begin{align}\label{curves}
\begin{cases}
1:\,k^{_{h}}=0\Rightarrow\chi_{_p}=\chi^{_-}_{_\alpha}-U_\uparrow+U_{_S},\\
2:\,\mathbf{k}_{_\|}=0\Rightarrow\chi_{_p}=\mu_\uparrow-U_\uparrow+E_{_\alpha},\\
3:\,k^{_{p}}=0\Rightarrow\chi_{_p}=\chi^{_+}_{_\alpha}-U_\uparrow+U_{_S},\\
4:\,k_{_{h}}=0\Rightarrow\chi_{_p}=2\varepsilon_{_\alpha}-U_\uparrow+\widetilde{m}U_\downarrow\\
\quad\quad\quad\quad\quad\quad\quad\quad\quad\quad-U_{_S}(\widetilde{m}-1).\\
\end{cases}
\end{align}
The curves $1$-$4$ separate the possible scattering regimes and
are shown in Figs.~\ref{fig2a} and~\ref{fig2b} where the latter
depicts a general (gerenic) diagram and the former assumes equal
species masses and zero HF potentials.

To better understand the various scattering regimes, we switch to
working with the energy $\varepsilon_{_\alpha}$ of
Eq.~\eqref{additional1} instead of $E_{_\alpha}$. Recall that
$\varepsilon_{_\alpha}$ takes values in the interval
$[\varepsilon_{_\alpha}^{_0},\infty[$ where
$\varepsilon_{_\alpha}^{_0}=[h_{_S}(1+\widetilde{m})+\mu_{_S}
(1-\widetilde{m})]/2$; clearly $\varepsilon_{_\alpha}=E_{_\alpha}$
in case $h_{_S}=0$ and $m_\uparrow=m_\downarrow$. If the energy of
the particle incident on the interface is small enough or
$\varepsilon_{_\alpha}<\sqrt{\widetilde{m}}\Delta$, there is
insufficient energy to excite quasiparticles inside the SF; the
particle is then fully reflected as a superposition of a particle
and a hole. This regime is labelled~I in Figs.~\ref{fig2a}
and~\ref{fig2b}. The situation where
$E_{_\alpha}<\chi_{_p}-\mu_\uparrow+U_\uparrow$ i.e.~regime II is
physically forbidden since the wave vector of the incident
particle (i.e.~$\mathbf{k}_{_\|}$) is imaginary.

Suppose now that a particle is incoming with energy
$\varepsilon_{_\alpha}$ above the threshold value
$\sqrt{\widetilde{m}}\Delta$. Although a particle in regime III
appears to have sufficient energy for transmission, in undergoes
complete reflection. Particles in regime IV may be transmitted and
reflected but all targeted states are particle-like. In both
regimes IV and V, there are no reflected holes; that is, Andreev
reflection does not occur. However, in regime V, both
particle-like and hole-like excitations are present in the SF.
Finally, in regime VI, both particle-like and hole-like
excitations, or both Andreev and normal reflection, are allowed.

The blocking of scattering regimes can be understood as being
caused by a too large angle of incidence at the interface, similar
to total internal reflection. While the critical angle is roughly
constant for optical total internal reflection, here the critical
angles of incidence depend on the energy of the incoming particle.
Generally, a particle of species $\uparrow$ has an angle of
incidence $\theta$ which satisfies
$\tan\theta=|\mathbf{k}_{_\parallel}|/k_{_p}$, or in terms of the
energy:
\begin{align}\label{angles}
\tan\left(\theta\right)=\left(\frac{\mu_\uparrow-U_\uparrow+E_{_\alpha}}{\chi_{_p}}-1\right)^{1/2}.
\end{align}
The critical angles for transmission, hole-like transmission and
Andreev reflection can then be obtained as a function of the
energy $E_{_\alpha}$ by substitution of $\chi_{_p}$ from
Eq.~\eqref{curves} into Eq.~\eqref{angles}.

To summarize, the lightly-shaded regimes in Figs.~\ref{fig2a}
and~\ref{fig2b} describe normal particles of species $\uparrow$
which undergo complete reflection, while the rightmost,
heavily-shaded regime is unphysical. \textit{Only incident
particles in the unshaded regimes may excite quasiparticles inside
the SF.} If we consider now holes of spin species $\downarrow$
which are incident on the interface from the N phase, they can
only be transmitted in a regime analogous to~VI in the diagram.
Furthermore, one can find now also the scattering diagram, applied
for particles of spin species $\downarrow$ and holes of spin
species $\uparrow$ incident on the interface and scattering to the
$\beta$ branch. The scattering diagram is then qualitatively
similar to that of Fig.~\ref{fig2a} and the curves $1$-$4$ are
given by Eqs.~\eqref{curves} after performing the transformations
$h_{_S}\rightarrow -h_{_S}$,
$\widetilde{m}\rightarrow1/\widetilde{m}$,
$\alpha\rightarrow\beta$ and $\uparrow\leftrightarrows
\downarrow$.

\subsection{Discussion of the Scattering Diagram}
At this stage, we can discuss the influence of the interaction
parameter $k_{_F}a$, the HF potentials and the species mass
asymmetry on the diagram of Fig.~\ref{fig2a} as they may strongly
affect the transport properties which will be calculated in the
following section.

\textit{The interaction regime ---} First of all, the
$\chi_{_p}$-$\varepsilon_{_\alpha}$ diagrams of the deep BCS
interaction regime and the unitary interaction regime have a
different topology. Assume for now that $m_\uparrow=m_\downarrow$
and that $U_{_{S}}=U_\uparrow=U_\downarrow=0$. In the deep BCS
regime, the relation $2\Delta\ll \Delta+\mu_{_S}$ implies that
line~$2$ in Fig.~\ref{fig2a} is shifted horizontally to much
higher values of $\chi_{_p}$ so that regime~VI becomes by far the
most important scattering regime. Therefore quasiparticle
reflections occur almost only via the Andreev mechanism. It is
known that this involves transport of particles, but no transport
of energy across the interface. In contrast, at unitarity where
$2\Delta> \Delta+\mu_{_S}$ (since $\Delta\approx 1.16\mu_{_S}$ as
derived from Eq.~\eqref{eqstate2} in case
$m_\uparrow=m_\downarrow$ and $U_{\uparrow,\downarrow,S}=0$),
phase VI does not even exist. This means that quasiparticles with
energy $\varepsilon_{_\alpha}$ above the threshold $\Delta$ cannot
undergo Andreev reflection but only normal (specular) reflection,
by which neither particles nor energy are carried across the
interface.

\textit{The HF potentials ---} Assume again that
$m_\uparrow=m_\downarrow$; the HF potentials $U_{_{S}}$,
$U_\uparrow$ and $U_\downarrow$ are seen to be present in the
expressions~\eqref{curves} which describe the curves $1$-$4$ of
Fig.~\ref{fig2a}. However, it follows from these equations that
\textit{the HF potentials induce only horizontal shifts of the
curves $1$-$4$ but do not affect the upper and the lower bounds of
regime I}. Whereas line $2$ is shifted to the right, the curves
$1$, $3$ and $4$ are always shifted to the left. This can be
deduced from the fact that $U_{_S}< U_\uparrow\leq 0$ and
$U_{_\downarrow}< U_\uparrow\leq 0$ since the interactions are
attractive and the density of the SF exceeds the density of the N
phase.

\textit{The species mass asymmetry ---} In the presence of a
species mass asymmetry $m_\uparrow\neq m_\downarrow$, regime I may
become either larger or smaller. This is due to a shift of its
lower bound $\varepsilon_{_\alpha}^{_0}=[
h_{_S}(1+\widetilde{m})+\mu_{_S} (1-\widetilde{m})]/2$ (obtained
by taking $E_{_\alpha}=0$ in Eq.~\eqref{additional1}) and a shift
of the upper bound
$\varepsilon_{_\alpha}=\sqrt{\widetilde{m}}\Delta$. It is seen
that both shifts have the same sign but can either increase or
decrease the surface area of regime I. Also, the species mass
asymmetry induces a shift of line $4$ as is seen from
Eq.~\eqref{curves}.

\subsection{Transmission Coefficients}
In order to study the transport properties in the next section, we
first seek to relate the scattering amplitudes $A$ and $B$ of
Eqs.~\eqref{general} to the transmission coefficients. The
amplitudes are determined by matching the quasiparticle wave
functions of Eq.~\eqref{general} and their derivatives at the
interface $z=0$~\cite{demers}. In its most general form, this
leads to the equations:
\begin{widetext}
\begin{align}
\left( \begin{array}{c}
A_{_{\mathbf{k},+}}^{_{p,n}} + A_{_{\mathbf{k},-}}^{_{p,n}}\\
B_{_{\mathbf{k},+}}^{_{h,n}}+ B_{_{\mathbf{k},-}}^{_{h,n}}\\
k_{_{p}}(A_{_{\mathbf{k},+}}^{_{p,n}} - A_{_{\mathbf{k},-}}^{_{p,n}})\\
 k_{_{h}}(B_{_{\mathbf{k},+}}^{_{h,n}}- B_{_{\mathbf{k},-}}^{_{h,n}})
\end{array} \right)=
\left( \begin{array}{c} A_{_{\mathbf{k},+}}^{_{p,s}}
+A_{_{\mathbf{k},+}}^{_{h,s}}
+A_{_{\mathbf{k},-}}^{_{p,s}} + A_{_{\mathbf{k},-}}^{_{h,s}}\\
\varsigma_{_p}B_{_{\mathbf{k},+}}^{_{p,s}}+\varsigma_{_h}B_{_{\mathbf{k},+}}^{_{h,s}}+
\varsigma_{_p}B_{_{\mathbf{k},-}}^{_{p,s}}+ \varsigma_{_h}B_{_{\mathbf{k},-}}^{_{h,s}}\\
k^{_p} (A_{_{\mathbf{k},+}}^{_{p,s}} -
A_{_{\mathbf{k},-}}^{_{p,s}}) + k^{_h}
(A_{_{\mathbf{k},+}}^{_{h,s}}
- A_{_{\mathbf{k},-}}^{_{h,s}})\\
k^{_p}\varsigma_{_p}(B_{_{\mathbf{k},+}}^{_{p,s}}
-B_{_{\mathbf{k},-}}^{_{p,s}}) +
k^{_h}\varsigma_{_h}(B_{_{\mathbf{k},+}}^{_{h,s}}
-B_{_{\mathbf{k},-}}^{_{h,s}})
\end{array} \right),
\end{align}
\end{widetext}
where we defined $\varsigma_{_{p,h}}\equiv
\chi_{_\alpha}^{_\pm}/\Delta$\label{varsigmadef}. As argued
before, depending on the various scattering regimes, one must set
some scattering amplitudes to zero. One finds:
\begin{align*}
\begin{cases}
\text{for regime VI: }&B_{_{\mathbf{k},-}}^{_{h,n}}=A^{_{p,s}}_{_{\mathbf{k},-}}=B^{_{h,s}}_{_{\mathbf{k},+}}=0,\\
\text{for regime V: }&B_{_{\mathbf{k},+}}^{_{h,n}}=A^{_{p,s}}_{_{\mathbf{k},-}}=B^{_{h,s}}_{_{\mathbf{k},+}}=0,\\
\text{for regime IV \& V: }& B_{_{\mathbf{k},+}}^{_{h,n}}=A^{_{p,s}}_{_{\mathbf{k},-}}=B^{_{h,s}}_{_{\mathbf{k},-}}=0,\\
\text{for holes:
}&A_{_{\mathbf{k},+}}^{_{p,n}}=A^{_{p,s}}_{_{\mathbf{k},-}}=B^{_{h,s}}_{_{\mathbf{k},+}}=0.
\end{cases}
\end{align*}
We give the solutions for the scattering amplitudes in
expressions~\eqref{app2}-\eqref{app5} in the Appendix.

From the HFBdG Eqs.~\eqref{Bdg} it readily follows that the
quasiparticle density
$\rho_{_{\alpha}}(\mathbf{r})=|u_\uparrow(\mathbf{r})|^2+|v_\downarrow(\mathbf{r})|^2$
and the quasiparticle currents

\begin{align*}
\mathbf{j}_{_{\alpha}}&=-\frac{i}{2m_\uparrow}\left[u_\uparrow^*\boldsymbol{\nabla}
u_\uparrow-u_\uparrow\boldsymbol{\nabla} u_\uparrow^*\right]
-\frac{i }{2m_\downarrow}\left[v_\downarrow\boldsymbol{\nabla}
v_\downarrow^*-v_\downarrow^*\boldsymbol{\nabla}
v_\downarrow\right],
\end{align*}
satisfy the ``continuity equation''
$\partial\rho_{_{\alpha}}/\partial
t+\boldsymbol{\nabla}\cdot\mathbf{j}_{_{\alpha}}=0$. Note,
however, that they do not express conservation of particles or
mass. The same expressions are valid for the $\beta$ branch after
the transformation $\uparrow\rightleftarrows\downarrow$. We define
now the \textit{transmission coefficient}
$S_{_\alpha}^{_p}(\varepsilon_{_\alpha},\chi_{_p})$\label{transdef}
of an incident particle of energy $\varepsilon_{_\alpha}$ and
momentum $\mathbf{k}$ as the ratio of the transmitted to the
incoming current along the $z$-axis. As argued before, the
transmission coefficients vanish in regimes~I, II and III of
Fig.~\ref{fig2a}; for regimes IV, ~V and~VI (and for holes) we
give their analytical forms in expression~\eqref{transmissions} in
the Appendix. For energies $\varepsilon_{_\alpha}$ slightly above
the transmission threshold $\varepsilon_{_\alpha}\approx
\sqrt{\widetilde{m}}\Delta$ (for regimes~V and VI) one can
analytically find the following behavior
\begin{align}\label{squarroot}
S_{_\alpha}^{_p}(\varepsilon_{_\alpha},\chi_{_p})\propto
\sqrt{\varepsilon_{_\alpha}-\sqrt{\widetilde{m}}\Delta}.
\end{align}
This is similar to the case of particles scattering from a HF
potential of height $\sqrt{\widetilde{m}}\Delta$~\cite{landau}.
Moreover, in the BCS regime wherein, by the Andreev approximation,
one can assume $k^{_{p}}=k^{_{h}}=k_{_{p}}=k_{_{h}}$, the
transmission coefficients reduce to:
\begin{align*}
S_{_\alpha}^{_{p,h}}&=\frac{2\sqrt{\varepsilon_{_\alpha}^2-\Delta^2\widetilde{m}}}{\varepsilon_{_\alpha}+\sqrt{\varepsilon_{_\alpha}^2-\Delta^2\widetilde{m}}}.
\end{align*}
Apart from the species mass asymmetry $\widetilde{m}$, this is the
expression which was obtained by Andreev in Ref.~\cite{andreev}.

\section{Heat Transport Through the Interface}\label{sec_transort} Thermally
excited particles incident on the surface will be reflected unless
they have an energy of at least
$2(\sqrt{\widetilde{m}}\Delta-\varepsilon_{_\alpha}^{_0})/(\widetilde{m}+1)$.
At low enough temperatures, not many particles will have enough
energy to penetrate. This leads to a significant decrease of heat
conductivity through the interface, a phenomenon well-known for
superconductors~\cite{andreev,andreev2}. We will now argue that
the same blocking of energy transport occurs for imbalanced
fermion gases, and discuss the influence of the interaction regime
on this phenomenon.

Heat transport across the interface can be studied using the
transmission coefficients calculated in last section. Using the
relation:
\begin{align*}
\int
\frac{\text{d}^3\mathbf{k}}{(2\pi)^3}=\frac{m_\uparrow}{2\pi^2(1+\widetilde{m})\hslash^2}\int
\text{d}\varepsilon_{_\alpha}\int \text{d}k_{_p},
\end{align*}
one can write the particle-like \textit{heat flux} through the
interface and to the $\alpha$ branch as:
\begin{align}\label{flux}
  W^{_p}_{_\alpha}=\frac{m_\uparrow}{\pi^2\hslash^3(\widetilde{m}+1)^2}&\int\text{d}\varepsilon_{_\alpha}\int
\text{d}\chi_{_\alpha}\\
&\times(\varepsilon_{_\alpha}-\varepsilon^{_0}_{_\alpha})f(E_{_\alpha}(\varepsilon_{_\alpha}))S_{_\alpha}^{_p}(\varepsilon_{_\alpha},\chi_{_\alpha}),\nonumber
\end{align}
where $f$\label{fermidef} is the Fermi-Dirac distribution and the
integration is performed over all regimes of the
$\chi_{_p}$-$\varepsilon_{_\alpha}$ plane. For the heat flux
$W^{_h}_{_\alpha}$ carried by \textit{hole} excitations, the
integration is only over regime VI and by use of
$\text{d}k_{_h}^2=\text{d}k_{_p}^2$, we can obtain the same
expression as Eq.~\eqref{flux} but with $m_\downarrow$ instead of
$m_\uparrow$ and with $S_{_\alpha}^{_h}$ instead of
$S_{_\alpha}^{_p}$, as given in~\eqref{transmissions4}. By the
same approach, the contributions of the $\beta$-branch are then
found by performing the transformations $h_{_S}\rightarrow
-h_{_S}$, $\widetilde{m}\rightarrow1/\widetilde{m}$,
$\alpha\rightarrow\beta$ and $\uparrow\leftrightarrows
\downarrow$.

\begin{figure}
\begin{center}
   \epsfig{figure=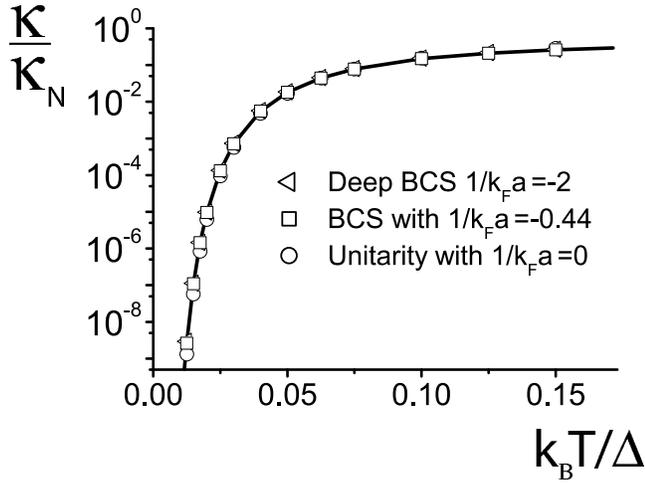,angle=0, height=180pt}
       \caption{The thermal conductivity $\kappa$ across the N-SF interface
         divided by the normal-phase
       conductivity $\kappa_{_N}$ against $k_{_B}T/\Delta$ for the unitary, BCS and deep
       BCS regimes when $m_\uparrow=m_\downarrow$ and $U_{\uparrow,\downarrow,S}=0$. For $k_{_B}T\lesssim 0.1\Delta$,
        $\kappa/\kappa_{_N}$ drops
        dramatically (notice the logarithmic scale).
        The curve represents the analytical result
        obtained using of the Andreev approximation, Eq.~\eqref{andreev}.
        At unitarity, $\Delta=0.69k_{_B}T_{_F}$, in the BCS case $\Delta=0.25 k_{_B}T_{_F}$
       and in the deep BCS case $\Delta=0.002 k_{_B}T_{_F}$. For
       large values of the temperature, one finds that
       $\kappa/\kappa_{_N}=1$.
       \label{fig3}
     }
\end{center}
\end{figure}

In equilibrium, quasiparticles are incident on both sides of the
interface such that the N-SF flux is balanced by an equal SF-N
flux. Suppose, however, that the temperature on the N side is
$\delta T$ higher than on the SF side (and $\delta T/T\ll 1$). A
net heat flow $\delta Q=\delta Q^{_\alpha}+\delta Q^{_\beta}$ then
results from the scattering to the $\alpha$-channel ($\delta
Q^{_\alpha}$) and to the $\beta$-channel ($\delta Q^{_\beta}$).
For small temperature differences, we may write $\delta
Q^{_\alpha}=\kappa^{_\alpha}\,\delta T$ with $\kappa^{_\alpha}$
the \textit{heat conductivity} caused by scattering to the
$\alpha$ channel. Within the Kapitza approach, we have:
\begin{align}\label{conductivity}
\kappa^{_\alpha}=\frac{\partial
(W_{_\alpha}^{_p}+W_{_\alpha}^{_h})}{\partial(k_{_B} T)}.
\end{align}
In the deep BCS regime, where the Andreev approximation may be
used, one may obtain an analytical expression for $\kappa$. Taking
$k^{_{p}}=k^{_{h}}=k_{_{p}}=k_{_{h}}$ and assuming the temperature
to satisfy $k_{_B}T\ll \Delta\ll \mu_{_S}$ effects regime VI of
Figs.~\ref{fig2a} and \ref{fig2b} to be the sole important
scattering regime; in other words, all reflections are due to
Andreev reflection. We obtain for the heat flux to the $\alpha$
channel (see Eq.~\eqref{flux}):
\begin{align*}
W_{_\alpha}^{_p}+&W_{_\alpha}^{_h}=\frac{\sqrt{\pi}
(m_\uparrow+m_\downarrow)\mu_\uparrow e^{-2\beta(\Delta\sqrt{\widetilde{m}}-\varepsilon^{_0}_{_\alpha})/(\widetilde{m}+1)}}{2\pi^2\hslash^3\sqrt{\widetilde{m}+1}}\\
&\times \frac{
[\Delta\sqrt{\widetilde{m}}-\varepsilon^{_0}_{_\alpha}+3k_{_B}T(\widetilde{m}+1)/4](k_{_B}T)^{3/2}}{\sqrt{\Delta\sqrt{\widetilde{m}}}}.
\end{align*}
For the $\alpha$-channel the heat conductivity of
Eq.~\eqref{conductivity} in the BCS regime is then obtained:
\begin{align}\label{andreev}
&\kappa_{_{BCS}}^{_\alpha}=\frac{\sqrt{\pi}
(m_\uparrow+m_\downarrow)e^{-2\beta(\Delta\sqrt{\widetilde{m}}-\varepsilon^{_0}_{_\alpha})/(\widetilde{m}+1)}}
{\pi^2\hslash^3(\widetilde{m}+1)^{3/2}}\\
&\times \frac{(\Delta\sqrt{\widetilde{m}}
-\varepsilon^{_0}_{_\alpha})[\Delta\sqrt{\widetilde{m}}
-\varepsilon^{_0}_{_\alpha}+3k_{_B}T(\widetilde{m}+1)/2]}{\sqrt{k_{_B}T\Delta\sqrt{\widetilde{m}}}},\nonumber
\end{align}
which amounts to Andreev's result when $h_{_S}=0$ and
$m_\uparrow=m_\downarrow$~\cite{andreev}. The heat flux and heat
conductivity of the $\beta$-channel is again found after the
transformation $\widetilde{m}\rightarrow 1/\widetilde{m}$,
$\uparrow \rightleftarrows\downarrow$, $h_{_S}\rightarrow-h_{_S}$
and $\alpha\rightarrow\beta$. The energy carried by the $\beta$
branch is a factor $e^{-4\beta h_{_S}/(\widetilde{m}+1)}$ lower
than that of the $\alpha$ branch and due to coexistence $h\sim
\Delta$ and since $\Delta\gg T$, it can be neglected.

Eq.~\eqref{andreev} demonstrates that $\kappa$ decays
exponentially fast with decreasing temperature. This can be
understood as follows. For low enough temperature
i.e.~$k_{_B}T\ll\sqrt{\widetilde{m}}\Delta$, only incident
particles (and holes) with energies slightly above the threshold
$\varepsilon_{_\alpha}\approx \sqrt{m}\Delta$ can contribute to
the heat conductivity; they have, however, a very low statistical
weight:
\begin{align}\label{FD}
f\propto
e^{-2\beta(\sqrt{\widetilde{m}}\Delta-\varepsilon_{_\alpha}^{_0})/(\widetilde{m}+1)}\ll
1.
\end{align}
It is this weight factor which also appears in
expression~\eqref{andreev}. The question now is: what will happen
to this exponential decay if we tune the interactions from the BCS
to the unitary regime? In order to answer this, we have
numerically calculated the ratio of $\kappa\equiv
\kappa^{_\alpha}+\kappa^{_\beta}$ to the conductivity in the N
phase $\kappa_{_N}= k_{_B}T(\mu_\uparrow m_\uparrow+\mu_\downarrow
m_\downarrow)/\pi^2\hslash^3$, and show this in Fig.~\ref{fig3} as
a function of $k_{_B}T/\Delta$ and we assumed
$m_\uparrow=m_\downarrow$ and $U_{\uparrow,\downarrow,S}=0$.
\textit{We find that $\kappa/\kappa_{_N}$ decreases drastically
below $k_{_B}T\approx 0.1 \Delta$ and is independent of the
interaction parameter $k_{_F} a$}. Also in Fig.~\ref{fig3} we plot
the heat conductivity of Eq.~\eqref{andreev}; it is seen to give a
good fit for all scattering regimes~\footnote{Note that $h_{_S}$
is eliminated using the coexistence condition
Eq.~\eqref{clogston}.}. The reason for this good agreement is that
the threshold line $\varepsilon_{_\alpha}=\sqrt{m}\Delta$ of the
diagram of Fig.~\ref{fig2a} is unaffected by changing the
scattering regime. Furthermore, relation~\eqref{squarroot} which
describes the transmission coefficient slightly above the
threshold line guarantees that $S_{_\alpha}^{_p}$ displays the
same behavior in regimes V~and~VI.

As argued before, the main effect of varying $1/(k_{_F} a)$ from
large and negative (deep BCS) to zero (unitarity), is to shift the
curves $1$-$4$ in diagram of Fig.~\ref{fig2a}. Both the strong
exponential decay in $\sqrt{\widetilde{m}}\Delta/(k_{_B}T)$ and
the square root dependence of $S_{_\alpha}^{_p}$ on
$\varepsilon_{_\alpha}-\sqrt{\widetilde{m}}\Delta$ are unaffected
by the variation of $1/(k_{_F} a)$ and the introduction of the
Hartree potentials, hence the very similar behavior of
$\kappa/\kappa_{_N}$ for all scattering regimes under study. On
the other hand, what is different between the scattering regimes
is the value of $\Delta$; it is small in the sense that $\Delta\ll
k_{_B}T_{_F}$ in the BCS limit and is comparable with
$k_{_B}T_{_F}$ at unitarity.

\begin{figure}
\begin{center}
   \epsfig{figure=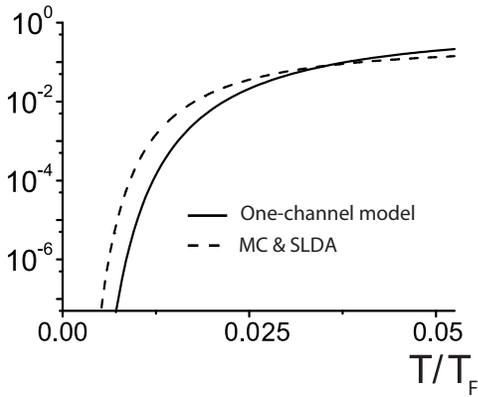,angle=0, height=150pt}
       \caption{The thermal conductivity $\kappa$ across the N-SF interface
         divided by the normal-phase
       conductivity $\kappa_{_N}$ against $T/T_{_F}$ at the interface in the unitary regime. The full line depict the Andreev approximation
wherein the coexistence condition is taken from the one-channel
model. The dotted line is the Andreev approximation with the
coexistence condition is taken from Monte Carlo (MC)
simulations~\cite{lobo} and a superfluid local-density
approximation (SLDA)~\cite{bulgac}. At the lowest experimental
values, that is when $T<0.03T_{_F}$, the conductivity is already
very low.\label{fig3bis}
     }
\end{center}
\end{figure}
\section{Equilibration Timescales}\label{sec_timescales} The introduction of the Hartree
potentials has two implications on the heat resistivity. First of
all, the scattering diagram will change by horizontal shifts of
lines $1$-$4$; as argued before, this does not affect the steep
exponential decay of resistivity. Secondly, the coexistence
condition will change, that is, the relation between $\Delta$ and
$h$ will change. This, on the other hand may strongly affect the
exponential decay as is seen from Eq.~\eqref{FD}. At unitarity,
the relation between $\Delta$ and $h$ is known from Monte Carlo
(MC)~\cite{lobo} simulations and a superfluid local density
approximation (SLDA)~\cite{bulgac}. From these we adapt the values
$\Delta=0.504\,k_{_B}T_{_F}$, $U_{_S}=-0.516\,k_{_B}T_{_F}$,
$\mu_{_S}=0.936\,k_{_B}T_{_F}$~\cite{bulgac} and finally
$h_{_S}=0.48(\mu_\uparrow+\mu_\downarrow)$~\cite{lobo}. In
Fig.~\ref{fig3bis}, we compare the Andreev approximation with
values of the one-channel model (full line) and of the MC and SLDA
results (dashed line). We see that the one-channel model predicts
the sharp decrease of $\kappa$ to be at higher temperatures than
the MC and SLDA results.

We find that the interface conductivity at the lowest experimental
temperatures is very low in the sense that the resulting
characteristic equilibration time is of order of seconds i.e. the
trap lifetime.

We explain now how we estimate the characteristic equilibration
time $\tau$.  Consider again a temperature bias $\delta T$ across
the interface. We are interested in the resulting total heat flux
$\delta W$, induced by the excess heat
$\sum_{i=\uparrow,\downarrow}\delta Q_i$ in the N phase. By
definition, the heat flux $W$ is the amount of heat transferred
per unit time per unit surface area. One can therefore estimate
$\tau$ by the relation $\delta
W=\sum_{i=\uparrow,\downarrow}\delta Q_i/(\tau \mathcal{A})$ with
$\mathcal{A}$ the interface area. Moreover, at position
$\mathbf{r}$ in the degenerate N phase $i=\uparrow,\downarrow$,
one has $\delta Q_i(\mathbf{r})=\pi^2\rho_i(\mathbf{r}) T\delta
T\text{d}\mathbf{r}/[2\mu_i(\mathbf{r})]$ with $\rho_i$ the
particle density and $i=\uparrow,\,\downarrow$~\cite{huang}.
Combining this with $\delta W=\kappa k_{_B}\delta T$, one finds
\begin{align*}
\tau=\frac{\pi^2k_{_B}
T}{2\mathcal{A}\kappa}\sum_{i=\uparrow,\downarrow}\int_N\frac{\rho_i(\mathbf{r})}{
\mu_i(\mathbf{r})}\text{d}\mathbf{r}.
\end{align*}
Assume now an isotropic trap of $10^7$ particles at polarization
$0.5$ with radial trapping frequency $458$Hz such that the local
density approximation is valid~\cite{zwierlein3}. For this
configuration, we calculate that the characteristic time of
relaxation $\tau$ is of order seconds when $T/T_{_F}\lesssim
0.02$. This means that according to our estimates, the relaxation
time is of the order of the lifetime of the system at the lowest
attainable experimental temperature $T/T_{_F}=0.03$ such that in
these experiments, the N and SF phases may not be equilibrated.

At least two interesting experimental consequences of the low
thermal conductivity for ultracold gases are different to what is
the case for a SC. First of all, in a superconductor, the heat
conductivity has a lattice component which dominates the electron
component at low temperatures; the lack of such a component in the
system under study makes the decrease of the conductivity more
significant~\cite{tinkham}. Secondly, at very low temperatures, a
N-SF temperature gradient will entail a pressure difference
$\delta P$ across the interface, and therefore the physical
movement of the interface; this is impossible for
normal-superconductor junctions. Taking $\delta P=(\partial
P/\partial T)\delta T$ and an infinite thermal resistivity, one
finds~\cite{huang}:
\begin{align}
\delta
P=\frac{5\pi^2}{6}\sum_{i=\uparrow,\,\downarrow}P_{_{i0}}\frac{k_{_B}T}{\mu_i}\times
\delta \left(\frac{k_{_B}T}{\mu_i}\right),
\end{align}
where $P_{_{i0}}$ is the $T=0$ pressure of species
$i=\uparrow,\downarrow$. Depending on the time scales involved,
mechanical equilibrium might be reached before thermal
equilibrium.

\section{Discussion}
The step function model for the gap (see Eq.~\eqref{gap}) is
adequate for tackling the problem at hand, that is for estimating
the heat conductivity across the interface.

First of all, this model should gives qualitatively correct
results as long as the quasiparticle wavelengths are of the same
order or less as the length over which the gap varies. Indeed,
this is clearly valid in the BCS limit, when the healing length
for the gap, further denoted as $\xi$, is of order
$\xi_{_0}=\hslash^2 k_{_F}/m_{_+}\Delta$ and $\Delta\ll
\hslash^2k_{_F}^2/m_{_+}$. Moreover, near unitarity, the
step-function model is also justified since the length of
variation of the gap is smaller than the quasiparticle
wavelength~\cite{desilva}. Therefore, the step function of
Eq.~\eqref{gap} serves as a good approximation for all regimes
under consideration.

Second, the step-model for the gap is justified whenever the
interface width must be small as compared to the variation length
of the trapping potential. This is also realized in experiments.

We argue now that, even if we would take into account the exact
interface profile, our results would hardly be affected. By
solving the Bogoliubov-de-Gennes equations in a self-consistent
manner, the interface profile for the gap could be
determined~\cite{mcmillan,tanaka}. At zero temperature, the
concomitant interface width is of order $\xi_{_0}=\hslash^2
k_{_F}/m_{_+}\Delta$. As was shown in Ref.~\onlinecite{vanson},
such smooth interface gives rise to a zero transmission
coefficients for particles with energy below the gap, and above
the gap, smaller transmission coefficients than predicted by our
step-function model. This simply means that particles incoming to
the interface with energy below the gap cannot cross over to the
superfluid phase and therefore they cannot carry energy across the
interface. Hence our conclusions of a blocking of energy transport
across the interface remains valid, independent of the interface
width. Note also that, due to the decreased transmission
coefficient above the gap, the heat conductivity will be even
smaller than predicted here.

The same conclusion can be drawn at finite temperatures: particles
with energy below the gap cannot penetrate the superfluid and the
conclusions of Ref.~\onlinecite{vanson} concerning the
transmission coefficients will still be valid. Indeed, since it is
known that the interface width increases with temperature as it
varies between $\xi(T=0)=\xi_{_0}$ and $\xi(T)\propto\xi_{_0}
(1-T/T_{_c})^{-1/2}$ near $T=T_{_c}$~\cite{degennes}, $\xi$ is
certainly of the order of $\xi_{_0}$ at the experimentally
relevant temperatures, that is below $0.1T_{_c}$. This can be seen
in Refs.~\onlinecite{kieselman} and~\onlinecite{nagato} where the
self-consistent calculations were performed for various
temperatures and even for the strong coupling
regime~\cite{kieselman}.

In our model, we do not take into account the temperature
dependence of the gap. However, this is well justified as the gap
$\Delta(T)$ depends very weakly on temperature at low $T$. In the
BCS limit it is well-known that $\Delta(T)$ goes as
$\Delta(T)-\Delta(T=0)\propto -\sqrt{T}\exp(-\Delta(0)/T)$ (see
for example in Ref.~\onlinecite{abrikosov}). Moreover, this weak
variation is also present in the strong coupling
limit~\cite{stajic} and can be understood within the
single-channel model from the appearance of the factor
$\exp(-\Delta/T)$ in the finite-temperature gap equation.

\section{Conclusion}

We have studied the effects of a N-SF interface in experiments on
trapped imbalanced fermion gases. We clarify the nature of the
possible processes that may occur when a particle is incident onto
the interface. Such particles will be thermally excited at any
finite temperature, and act as carriers of energy, thus eventually
causing thermal equilibration between the N and SF phases.

We find that reflection (both Andreev and normal, or specular) of
low energy particles off the interface in experiments on trapped
imbalanced fermion gases causes a suppression of energy transport
from one phase to the other. At low enough temperatures
(comparable to the currently accessible temperatures in
experiments), this suppression grows exponentially with decreasing
temperature. This, in turn, delays thermal equilibration of the
system, which may result in a temperature difference between the N
and the SF. Our estimates of the timescales for this equilibration
are in the seconds, which indicates that this effect is
experimentally relevant.

The incorporation of this temperature difference may allow current
models to finally resolve the difficulties in explaining the
experiments.

\section{Acknowledgement}
It is a pleasure to thank Henk Stoof and Koos Gubbels for
discussions and useful suggestions. We acknowledge partial support
by Project No.~FWO G.0115.06; A.L.~is supported by Project
No.~GOA/2004/02 and B.V.S. by the Research Fund K.U.Leuven.

\section*{Appendix}
Assume particles of species $\uparrow$ incident on the interface
with momentum $k_{_p}$ and energy $\varepsilon_{_\alpha}$. Define
first $\varsigma_{_{p,h}}\equiv \chi_{_\alpha}^{_\pm}/\Delta$ and:
\begin{subequations}
\begin{align*}
\mathcal{J}_{_0}&\equiv\varsigma_{_p}(k^{_h}-k_{_p})(k^{_p}-k_{_h})-\varsigma_{_h}(k^{_h}+k_{_h})(k^{_p} + k_{_p}),\\
\mathcal{J}_{_1}&\equiv\varsigma_{_p}(k^{_h}-k_{_p})(k^{_p}+k_{_h})-\varsigma_{_h}(k^{_h}-k_{_h})(k^{_p}+k_{_p}),\\
\mathcal{J}_{_2}&\equiv\varsigma_{_p}(-k^{_h}-k_{_p})(k^{_p}+k_{_h})-\varsigma_{_h}(-k^{_h}-k_{_h})(k^{_p}+k_{_p}).
\end{align*}
\end{subequations}
If the coordinate $(\chi_{_p},\varepsilon_{_\alpha})$ with
$\chi_{_p}\equiv\hslash^2k_{_p}^2/(2m_\uparrow)$ is positioned in
regime VI of Fig.~\ref{fig2a}, the matching of the wave functions
and their derivatives at $z=0$ leads to the coefficients:
\begin{align}\label{app2}
 \left( \begin{array}{c}
A_{_{\mathbf{k},-}}^{_{p,n}}\\
\\
B_{_{\mathbf{k},+}}^{_{h,n}}\\
A_{_{\mathbf{k},+}}^{_{p,s}}\\
B_{_{\mathbf{k},-}}^{_{h,s}}
\end{array} \right)=\frac{A_{_{\mathbf{k},+}}^{_{p,n}}}{\mathcal{J}_{_0}}
\left( \begin{array}{c}
\varsigma_{_p}(k^{_h}+k_{_p})(k_{_h} -k^{_p})\\
\quad-\varsigma_{_h}(k^{_h}+k_{_h})(k_{_p} - k^{_p})\\
-2\widetilde{m}k_{_p}(k^{_h} +k^{_p})\\
-2\varsigma_{_h}k_{_p}(k^{_h} +k_{_h})\\
-2\varsigma_{_p}k_{_p}(k^{_p} -k_{_h})
\end{array} \right).
\end{align}
For regime V (see Fig.~\ref{fig2a}), we get:
\begin{align}\label{app3}
 \left( \begin{array}{c}
A_{_{\mathbf{k},-}}^{_{p,n}}\\
\\
B_{_{\mathbf{k},-}}^{_{h,n}}\\
A_{_{\mathbf{k},+}}^{_{p,s}}\\
B_{_{\mathbf{k},-}}^{_{h,s}}
\end{array} \right)=\frac{A_{_{\mathbf{k},+}}^{_{p,n}}}{\mathcal{J}_{_1}}
  \left( \begin{array}{c}
  \varsigma_{_p}(k^{_h}+k_{_p})(-k^{_p}-k_{_h})\\
 \quad-\varsigma_{_h}(k^{_h}-k_{_h})(k_{_p}-k^{_p})\\
-2\widetilde{m}k_{_p}(k^{_h} +k^{_p})\\
-2\varsigma_{_h}k_{_p}(k^{_h} -k_{_h})\\
-2\varsigma_{_p}k_{_p}(k^{_p} +k_{_h})
\end{array} \right).
\end{align}
For regime IV (see Fig.~\ref{fig2a}), we get:
\begin{align}\label{app4}
 \left( \begin{array}{c}
A_{_{\mathbf{k},-}}^{_{p,n}}\\
\\
B_{_{\mathbf{k},-}}^{_{h,n}}\\
A_{_{\mathbf{k},+}}^{_{p,s}}\\
B_{_{\mathbf{k},+}}^{_{h,s}}
\end{array} \right)=\frac{A_{_{\mathbf{k},+}}^{_{p,n}}}{\mathcal{J}_{_2}}
  \left( \begin{array}{c}
\varsigma_{_p}(-k^{_h}+k_{_p})(-k^{_p}-k_{_h})\\
\quad-\varsigma_{_h}(-k^{_h}-k_{_h})(k_{_p}-k^{_p})\\
-2\widetilde{m}k_{_p}(-k^{_h} +k^{_p})\\
-2\varsigma_{_h}k_{_p}(-k^{_h} -k_{_h})\\
-2\varsigma_{_p}k_{_p}(k^{_p} +k_{_h})
\end{array} \right).
\end{align}
For holes of species $\downarrow$ which are incident on the
interface, and which necessarily must have energy and momentum in
regime VI, the scattering amplitudes are:
\begin{align}\label{app5}
 \left( \begin{array}{c}
B_{_{\mathbf{k},+}}^{_{h,n}}\\
\\
A_{_{\mathbf{k},-}}^{_{p,n}}\\
A_{_{\mathbf{k},+}}^{_{p,s}}\\
B_{_{\mathbf{k},-}}^{_{h,s}}
\end{array} \right)=\frac{B_{_{\mathbf{k},-}}^{_{h,n}}}{\mathcal{J}_{_0}}
  \left( \begin{array}{c}
  \varsigma_{_h}(k^{_h}-k_{_h})(k^{_p}+k_{_p})\\
  \quad-\varsigma_{_p}(k^{_h}-k_{_p})(k^{_p} +k_{_h})\\
-2k_{_h}(k^{_h} +k^{_p})\\
-2k_{_h}(k^{_h} -k_{_p})\\
-2k_{_h}(k^{_p} +k_{_p})
\end{array} \right).
\end{align}
The transmission coefficients for the different regimes are:
\begin{subequations}\label{transmissions}
\begin{align}
\text{For regime VI: }S_{_\alpha}^{_p}=&\left[4k^{_p}k_{_p}(k^{_h}+k_{_h})^2(\varsigma_{_h}^2-\widetilde{m})\right.\label{transmissions1}\\
&\left.+4k^{_h}k_{_p}(k_{_h}-k^{_p})^2(\widetilde{m}-\varsigma_{_p}^2)\right]/\mathcal{J}_{_0}^2.\nonumber\\
\text{For regime V: }S_{_\alpha}^{_p}=&\left[4k^{_p}k_{_p}|k^{_h}-k_{_h}|^2(\varsigma_{_h}^2-\widetilde{m})\right.\label{transmissions2}\\
&\left.+4k^{_h}k_{_p}|k_{_h} +k^{_p}|^2(\widetilde{m}-\varsigma_{_p}^2)\right]/|\mathcal{J}_{_1}|^2.\nonumber\\
\text{For regime IV: }S_{_\alpha}^{_p}=&\frac{4k^{_p}k_{_p}|k^{_h}+k_{_h}|^2(\varsigma_{_h}^2-\widetilde{m})}{|\mathcal{J}_{_2}|^2}.\label{transmissions3}\\
\text{For holes: }
S_{_\alpha}^{_h}=&\left[4k^{_p}k_{_h}(k^{_h}-k_{_p})^2(\widetilde{m}-\varsigma_{_p}^2)\right.\label{transmissions4}\\
&\left.+4k^{_h}k_{_h}(k^{_p}+k_{_p})^2(\varsigma_{_h}^2-\widetilde{m})\right]/\mathcal{J}_{_0}^2.\nonumber
\end{align}
\end{subequations}
Here we also used $\varsigma_{_{h}}=\widetilde{m}/\varsigma_{_p}$
and $\varsigma_{_p}^2<\widetilde{m}<\varsigma_{_h}^2$.

\end{document}